\documentclass{article}

 \usepackage[main,final]{neurips_2026}
\usepackage{hyperref}       %

\usepackage{amsmath,amsfonts,bm}

\def\eqref#1{equation~\ref{#1}}

\def\1{\bm{1}}

\def\vb{{\bm{b}}}
\def\vc{{\bm{c}}}

\def\vh{{\bm{h}}}

\def\vu{{\bm{u}}}

\def\mM{{\bm{M}}}

\def\mW{{\bm{W}}}

\DeclareMathAlphabet{\mathsfit}{\encodingdefault}{\sfdefault}{m}{sl}
\SetMathAlphabet{\mathsfit}{bold}{\encodingdefault}{\sfdefault}{bx}{n}

\def\gC{{\mathcal{C}}}

\def\gF{{\mathcal{F}}}

\newcommand{\Ls}{\mathcal{L}}
\newcommand{\R}{\mathbb{R}}

\newcommand{\reg}{\lambda_{\mathrm{corr}}}

\newcommand{\Cov}{\mathrm{Cov}}

\usepackage[utf8]{inputenc} %
\usepackage{lipsum} %

\usepackage[T1]{fontenc}

\usepackage{etoolbox}
\newbool{includeappendix}
\setbool{includeappendix}{true}

\ifdefined\isoverfull
\else
\fi

\usepackage{subcaption}

\usepackage{xcolor} %

\definecolor{my-full-blue}{HTML}{1F77B4}

\definecolor{my-full-orange}{HTML}{FF7F0E}

\definecolor{my-full-green}{HTML}{2CA02C}

\definecolor{my-full-red}{HTML}{d62728}

\definecolor{my-full-purple}{HTML}{9467bd}

\definecolor{my-full-brown}{HTML}{8c564b}

\definecolor{my-full-pink}{HTML}{e377c2}

\definecolor{my-full-gray}{HTML}{7f7f7f}

\definecolor{my-full-olive}{HTML}{bcbd22}

\definecolor{my-full-cyan}{HTML}{17becf}

\definecolor{c1}{RGB}{86, 100, 26}
\definecolor{c2}{RGB}{192, 175, 251}
\definecolor{c3}{RGB}{230, 161, 118}
\definecolor{c4}{RGB}{0, 103, 138}
\definecolor{c5}{RGB}{152, 68, 100}
\definecolor{c6}{RGB}{94, 204, 171}
\definecolor{c7}{RGB}{205, 205, 205}

\definecolor{csabr}{RGB}{152, 68, 100}
\definecolor{cibp}{RGB}{94, 204, 171}
\definecolor{ctaps}{RGB}{0, 103, 138}

\definecolor{cm1}{HTML}{1f77b4}
\definecolor{cm2}{HTML}{ff7f0e}
\definecolor{cm3}{HTML}{2ca02c}
\definecolor{cm4}{HTML}{d62728}
\definecolor{cm5}{HTML}{9467bd}
\definecolor{cm6}{HTML}{8c564b}
\definecolor{cm7}{HTML}{e377c2}
\definecolor{cm8}{HTML}{7f7f7f}
\definecolor{cm9}{HTML}{bcbd22}
\definecolor{cm10}{HTML}{17becf}

\colorlet{my-blue}{my-full-blue!30}
\colorlet{my-orange}{my-full-orange!30}
\colorlet{my-green}{my-full-green!30}
\colorlet{my-red}{my-full-red!30}
\colorlet{my-purple}{my-full-purple!30}
\colorlet{my-brown}{my-full-brown!30}
\colorlet{my-pink}{my-full-pink!30}
\colorlet{my-gray}{my-full-gray!30}
\colorlet{my-olive}{my-full-olive!30}
\colorlet{my-cyan}{my-full-cyan!30}

\usepackage{listings}

\usepackage{textcomp}

\usepackage{xcolor}

\usepackage[scaled=0.8]{beramono}

\definecolor{ckeyword}{HTML}{7F0055}
\definecolor{ccomment}{HTML}{3F7F5F}
\definecolor{cstring}{HTML}{2A0099}

\lstdefinestyle{numbers}{
	numbers=left,
	framexleftmargin=20pt,
	numberstyle=\tiny,
	firstnumber=auto,
	numbersep=1em,
	xleftmargin=2em
}

\lstdefinestyle{layout}{
	frame=none,
	captionpos=b,
}

\lstdefinestyle{comment-style}{
	morecomment=[l]//,
	morecomment=[s]{/*}{*/},
	commentstyle={\color{ccomment}\itshape},
}

\lstdefinestyle{string-style}{
	morestring=[b]",%
	morestring=[b]',%
	stringstyle={\color{cstring}},
	showstringspaces=false,%
}

\lstdefinestyle{keyword-style}{
	keywordstyle={\ttfamily\bfseries},
	morekeywords={
		function,
		constructor,
		int,
		bool,
		return,
		returns,
		uint
	},
	morekeywords = [2]{},
	keywordstyle = [2]{\text},
	sensitive=true,
}

\lstdefinestyle{input-encoding}{
	inputencoding=utf8,
	extendedchars=true,
	literate=
	{ℝ}{$\reals$}1%
	{→}{$\rightarrow$}1%
	{α}{$\alpha$}1%
	{β}{$\beta$}1%
	{λ}{$\lambda$}1%
	{θ}{$\theta$}1%
	{ϕ}{$\phi$}1%
}

\lstdefinestyle{escaping}{
	moredelim={**[is][\color{blue}]{\%}{\%}},
	escapechar=|,
	mathescape=true
}

\lstdefinestyle{default-style}{
	basicstyle=\fontencoding{T1}\ttfamily\footnotesize,
	style=numbers,
	style=layout,
	style=comment-style,
	style=string-style,
	style=keyword-style,
	style=input-encoding,
	style=escaping,
	tabsize=2,
	upquote=true
}

\lstdefinelanguage{BASIC}{
	language=C++,
	style=default-style
}[keywords,comments,strings]%

\usepackage[capitalize,noabbrev]{cleveref}

\crefname{listing}{Lst.}{listings}
\crefname{line}{Lin.}{Lin.}
\crefname{appendix}{App.}{App.}
\crefname{lemma}{Lemma}{Lemmas}
\Crefname{lemma}{Lemma}{Lemmas}
\crefname{thm}{Theorem}{Theorems}
\Crefname{thm}{Theorem}{Theorems}

\newcommand{\app}[1]{%
	\ifbool{includeappendix}{\cref{#1}}{the appendix}%
}
\newcommand{\App}[1]{%
	\ifbool{includeappendix}{\cref{#1}}{The appendix}%
}

\usepackage{tikz}

\usetikzlibrary{calc,decorations,decorations.pathmorphing}
\usetikzlibrary{positioning,fit,arrows}
\usetikzlibrary{decorations.markings}
\usetikzlibrary{shapes,shapes.geometric}
\usetikzlibrary{shadows,patterns,snakes}
\usetikzlibrary{backgrounds,decorations.pathreplacing,automata}
\usetikzlibrary{intersections}
\usetikzlibrary{angles,quotes}
\usetikzlibrary{plotmarks}
\usetikzlibrary{patterns}
\usetikzlibrary{arrows.meta}
\usetikzlibrary{shapes.misc}
\usetikzlibrary{chains}
\usetikzlibrary{calc}

\usepackage[utf8]{inputenc} %
\usepackage[T1]{fontenc}    %
\usepackage{url}            %
\usepackage{booktabs}       %
\usepackage{amsfonts}       %
\usepackage{nicefrac}       %
\usepackage{microtype}      %
\usepackage{xcolor}         %

\usepackage{graphicx}
\usepackage{adjustbox}
\usepackage{threeparttable}
\usepackage{thmtools, nameref}
\usepackage{multirow}
\usepackage{amssymb}
\usepackage{pifont}
\usepackage{mathtools}
\usepackage{amsthm}
\usepackage[abbreviations]{foreign}
\usepackage{enumitem}
\usepackage{makecell}
\usepackage{algorithm}
\usepackage[noend]{algpseudocode} %
\usepackage{wrapfig}
\usepackage{float}

\newcommand{\cmark}{\textcolor[rgb]{0.12,0.56,0.24}{\ding{51}}}
\newcommand{\xmark}{\textcolor[rgb]{0.78,0.20,0.24}{\ding{55}}}

\crefname{lemma}{Lemma}{lemmas}
\Crefname{lemma}{Lemma}{Lemmas}
\crefname{corollary}{corollary}{corollaries}
\Crefname{corollary}{Corollary}{Corollaries}

\usepackage{siunitx}
\newcolumntype{d}[1]{S[table-format=#1]}

\colorlet{cbackground}{c7!20}
\colorlet{cexact}{my-full-blue!45}
\colorlet{cexactlatent}{my-full-green!40}
\colorlet{cfwd}{black!100}
\colorlet{cpgd}{my-full-purple!75}
\colorlet{cpgdsignle}{my-full-red!90!black!65}

\colorlet{cbwd}{my-full-red!90!black!65}
\colorlet{cibp}{black!80}
\colorlet{ctool}{c4!100}
\colorlet{netinside}{c7!100}

\tikzstyle{toolstyle}=[dotted, line width = 1.2pt,draw=ctool]
\tikzstyle{ibpstyle}=[dash pattern=on 5pt off 2pt, cibp]
\tikzstyle{exactstyle}=[fill=cexact, opacity=1.0, draw=none]
\tikzstyle{exactlatentstyle}=[fill=cexactlatent, opacity=0.8, draw=none]
\tikzstyle{point}=[line width=0.5pt, draw=black, cross out, inner sep=0pt, minimum width=3pt, minimum height=3pt, anchor=center]
\tikzstyle{wcpoint}=[line width=0.5pt, draw=black, fill=black, circle, inner sep=0pt, minimum width=2.5pt, minimum height=2.5pt, anchor=center]
\tikzstyle{pgdarrow}=[color=cpgd, thick]
\tikzstyle{pgdsinglearrow}=[color=cpgdsignle, thick]
\tikzstyle{fwdarrow}=[color=cfwd, thick, dashed]
\tikzstyle{bwdarrow}=[color=cbwd,  line width = 1.2pt,, dotted]
\tikzstyle{pane}=[fill=cbackground, rectangle, rounded corners=2pt]

\newcommand{\newprotectedcommand}[2]{\newcommand{#1}{\protecting{#2}}}

\newprotectedcommand{\markerfwd}{\tikz[]{\draw[-stealth, fwdarrow] (0, 0) -- (0.3, 0); \node[anchor=center, minimum width=0pt, minimum height=4pt, inner sep=0pt] () at (0.15, 0.0) {};}\xspace}
\newprotectedcommand{\markerbwd}{\tikz[]{\draw[-stealth, bwdarrow] (0, 0) -- (0.3, 0); \node[anchor=center, minimum width=0pt, minimum height=4pt, inner sep=0pt] () at (0.15, 0.0) {};}\xspace}
\newprotectedcommand{\markerpgd}{\tikz[]{\draw[-stealth, pgdarrow] (0, 0) -- (0.3, 0); \node[anchor=center, minimum width=0pt, minimum height=4pt, inner sep=0pt] () at (0.15, 0.0) {};}\xspace}
\newprotectedcommand{\markerexact}{\tikz[]{\node[fill, aspect=1, inner sep=0pt, minimum size=2.1mm, exactstyle]{};}\xspace}
\newprotectedcommand{\markerexactlatent}{\tikz[]{\node[fill, aspect=1, inner sep=0pt, minimum size=2.1mm, exactlatentstyle]{};}\xspace}
\newprotectedcommand{\markeribp}{\tikz[]{\node[aspect=1, draw=cibp, inner sep=0pt, minimum size=2.1mm, ibpstyle, dash pattern=on 2.5pt off 1pt]{};}\xspace}
\newprotectedcommand{\markertool}{\tikz[]{\draw[draw=ctool, toolstyle, dash pattern=on 1pt off 0.75pt] (0,0.2) -- (0.2,0.2) -- (0.2,0);}\xspace}
\newprotectedcommand{\markerwc}{\tikz[]{\node[wcpoint] at (0,0.01) {};\node[] at (0,0) {};}\xspace}
\newprotectedcommand{\markerpoint}{\tikz[]{\node[point]{};}\xspace}
\newprotectedcommand{\markersinglepgd}{\tikz[]{\draw[pgdsinglearrow, dashed, line width =1.1pt] (0, 0) -- (0.3, 0); \node[anchor=center, minimum width=0pt, minimum height=4pt, inner sep=0pt] () at (0.15, 0.0) {};}\xspace}

\newprotectedcommand{\markermultipgd}{\tikz[]{\draw[pgdarrow, dotted, dash pattern=on 1pt off 0.75pt] (0,0.2) -- (0.2,0.2) -- (0.2,0);}\xspace}

\newprotectedcommand{\markerwcsingle}{\tikz[]{\node[point,pgdsinglearrow] at (0,0.0) {};\node[] at (0,0) {};}\xspace}
\newprotectedcommand{\markeradvpoint}{\tikz[]{\node[point,cpgd]{};}\xspace}

\newprotectedcommand{\markersabr}{\tikz[]{\node[fill, diamond, color=csabr,inner sep=0,  minimum size=2.5mm, aspect=0.5]{};}\xspace}
\newprotectedcommand{\markertaps}{\tikz[]{\node[aspect=1, fill=c4, inner sep=0pt, minimum size=2.1mm]{};}\xspace}

\title{End-to-End Neural Modeling of EM Response and Design Performance for Free-Form RFIC Passives}

\author{%
  Yuhao Mao\textsuperscript{1}\thanks{Equal contribution.}\quad
  Chenhao Chu\textsuperscript{2,*}\quad
  Martin Vechev\textsuperscript{1}\quad
  Hua Wang\textsuperscript{2}\\
  \textsuperscript{1}Department of Computer Science\\
  \textsuperscript{2}Department of Information Technology and Electrical Engineering\\
  ETH Zurich, Switzerland\\
  \texttt{\{yuhao.mao,martin.vechev\}@inf.ethz.ch}\\
  \texttt{\{chenhao.chu,hua.wang\}@iis.ee.ethz.ch}
}

\begin{document}
\maketitle

\begin{abstract}
Passive matching networks are critical components in radio-frequency integrated circuits (RFICs), but their design is bottlenecked by expensive finite-difference-based electromagnetic (EM) simulation. We propose an end-to-end neural framework for free-form RFIC passives that predicts both EM response and downstream design performance under varying frequency and circuit contexts, combining a frequency-conditioned neural simulator with a circuit-conditioned neural solver. To improve design performance prediction, the simulator outputs a neural EM response consisting of both the explicit EM response and a complementary response, together with a dependence regularizer that encourages the two responses to capture orthogonal information. To evaluate the proposed framework, we construct a large-scale dataset of free-form passive layouts, synthesized from both expert-designed contours and random paths. Our 21M-parameter model achieves high accuracy across a broad frequency range, circuit contexts and passive matching settings, with $R^2>0.98$ for almost all targets, while providing orders-of-magnitude faster evaluation than traditional EM simulation. Code available at \url{https://github.com/ETH-IDEAS/RFIC-NeuralSim}.

\end{abstract}

\section{Introduction}
\label{sec:introduction}

Radio-frequency integrated circuits (RFICs) are core building blocks of modern wireless systems. Within an RFIC, the power amplifier (PA) is a central component of the transmit chain and has therefore received sustained design attention.
A key challenge in PA design is the passive matching network\footnote{In this paper, ``network'' refers to a passive matching network rather than a neural network unless stated otherwise.}. Its layout geometry determines the electromagnetic (EM) response, which in turn shapes downstream design performance.
Traditional finite-difference-based EM simulation is extremely expensive, preventing fast design optimization for RFIC passives. This cost has motivated fast surrogate models for RFIC passives, including recent neural approaches.

Existing neural approaches have made progress on RFIC passive modeling, but their coverage remains limited, as summarized in \cref{tab:survey}.
First, many prior methods operate on templated or otherwise restricted passive layouts, which simplifies modeling but significantly narrows the design space.
Second, most prior work focuses on EM response prediction without modeling downstream design performance.
This is a serious but overlooked limitation because small EM-response errors can cause large design-performance degradations when passed to a ground-truth solver, as shown in \cref{fig:solver_instability}.
Third, prior methods often omit explicit auxiliary circuit conditioning, operate on narrow or sparse frequency regimes, or require retraining when the frequency grid changes.
These gaps leave open the central problem we study here: end-to-end neural modeling of free-form passive matching networks, from layout to both EM response and design performance, under varying circuit and frequency settings.

\begingroup
\setlength{\intextsep}{4pt}
\begin{wrapfigure}{r}{0.46\textwidth}
    \centering
    \includegraphics[width=\linewidth]{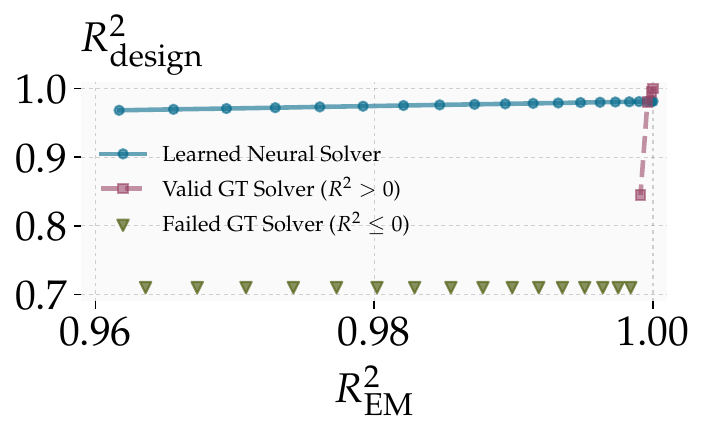}
    \captionsetup{font=small}
    \caption{Solver stability under imperfect EM response. The ground-truth (GT) solver achieves $R^2=1$ with noiseless input, but its design-performance $R^2$ becomes negative even at $R^2_{\mathrm{EM}}\approx 0.9984$. In contrast, the learned neural solver remains stable given imperfect EM responses. \Cref{app:solver_stability} details the evaluation protocol and illustrates the magnitude of the perturbations.}
    \label{fig:solver_instability}
\end{wrapfigure}

Here, \emph{design performance} refers to the downstream circuit quantities induced by the passive matching network within its surrounding circuit context, such as auxiliary capacitors.

In this work, we address these limitations with an end-to-end model for free-form passive matching networks under varying circuit and frequency settings.
Our model combines a frequency-conditioned \emph{neural simulator} with a circuit-conditioned \emph{neural solver}, as shown in \cref{fig:forward_model_arch}.
The neural simulator maps passive layout and frequency to a \emph{neural EM response}, which consists of the explicit EM response and a complementary response.
The complementary response preserves information that is useful for design-performance prediction when the explicit EM response is imperfect.
The neural solver then combines the neural EM response with circuit context to predict design performance.

Our main contributions are as follows:

\WFclear
\endgroup
\begin{itemize}[leftmargin=1.5em]
    \item We introduce an end-to-end neural modeling framework for free-form RFIC passives.
    Two components are proposed to further improve this framework: a complementary response that avoids the information bottleneck of the imperfect EM response prediction, and a dependence regularizer that encourages independence between the EM response and complementary response.
    \item We construct a large-scale dataset of free-form RFIC passives, with layouts synthesized from both expert-designed contours and random paths, and use it for a comprehensive empirical study.
    \item Our 21M-parameter model achieves high accuracy ($R^2 > 0.98$ for almost all targets) across a broad frequency range. On an RTX 2080 Ti GPU, the neural simulator runs at 121 full-frequency simulations per second, and the neural solver runs at 25{,}069 full-frequency predictions per second. Compared with traditional EM simulation, which can take up to 20 minutes for one full-frequency evaluation, our method provides orders-of-magnitude acceleration while maintaining high accuracy, enabling fast design optimization for RFIC passives.
\end{itemize}
Aside from the closest prior works in \cref{tab:survey}, a more detailed related work can be found in \cref{sec:related_work}.

\begin{table}
\centering
\caption{Comparison between our method and prior work on neural modeling of RFIC passives.}
\label{tab:survey}
\begin{adjustbox}{max width=\linewidth,center}

\begin{threeparttable}
\begin{tabular}{ccccccc}
\toprule
\multirow{3}{*}{} & \multirow{3}{*}{\makecell{Free-form \\ Passive Layout}} & \multicolumn{2}{c}{Modeling Outcome} & \multicolumn{2}{c}{\makecell{Modeling Conditions}} & \multirow{3}{*}{\makecell{Open\\Source}} \\
\cmidrule(lr){3-4}\cmidrule(lr){5-6}
 &  & \makecell{EM\\Response} & \makecell{Design\\Performance} & \makecell{Frequency} & \makecell{Circuit\\Context} & \\
\midrule
\citet{yan2018multi} & \xmark & \cmark & \xmark & 100 pts, 1 GHz range & not applicable & \xmark \\
\citet{guan2021modeling} & \xmark & \cmark & \xmark & 20 pts, 20 GHz range & not applicable & \xmark \\
\citet{li2022deep} & \xmark & \xmark & \cmark & \xmark & \xmark & \xmark \\
\citet{lee2022neural} & \xmark & \cmark & \cmark & 10 pts, 100 GHz range & \xmark & \xmark \\
\citet{karahan2024deep} & \cmark & \cmark & \xmark & pts not reported, 76 GHz range$^\dagger$ & not applicable & \xmark \\
\citet{sahu2025surrogate} & \xmark & \cmark & \xmark & pts not reported, 2 GHz range$^\dagger$ & not applicable & \xmark \\
\citet{zhang2026efficient} & \xmark & \cmark & \xmark & pts not reported, 20 GHz range & not applicable & \xmark \\
Ours & \cmark & \cmark & \cmark & 501 pts, 50 GHz range & \cmark & \cmark \\
\bottomrule
\end{tabular}
\begin{tablenotes}[flushleft]
\footnotesize
\item $^\dagger$ Instead of explicit conditioning, they use separate output dimensions for different frequencies, which requires retraining the model to operate on unseen frequencies.
\end{tablenotes}
\end{threeparttable}
\end{adjustbox}
\end{table}

\section{Background}
\label{sec:background}

\begin{figure}
    \centering
    \includegraphics[width=0.98\linewidth]{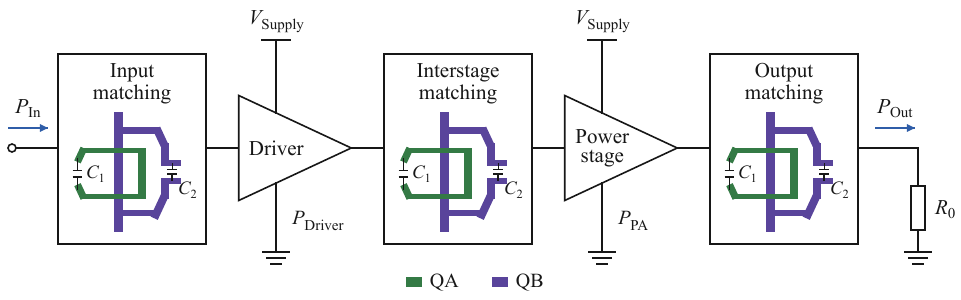}
    \caption{Typical RFIC PA with three passive matching networks: input matching, interstage matching, and output matching.}
    \label{fig:pa_block_diagram}
\end{figure}

A typical RFIC power amplifier (PA) uses passive matching networks at its input, between stages, and at its output, as shown in \cref{fig:pa_block_diagram}. Informally, these networks are small on-chip passive circuits that reshape how RF signals pass between blocks. Since this behavior is set by the metal geometry, the layout affects both the local electromagnetic (EM) response and the end-to-end PA performance.

We study a realistic passive layout setting. Each passive layout is symmetric, its ports are mutually connected, and its geometry is represented on a fixed $18\times18$ pixel grid. The layout contains two metal layers, \emph{QA} and \emph{QB}, shown as the green and purple layers in \cref{fig:pa_block_diagram}. The chosen passive structures are reciprocal because their EM model uses linear, time-invariant conductors and dielectrics with reciprocal material responses. The surrounding circuit context is parameterized by two scalar capacitor values, $C_1$ and $C_2$. The input variables for the prediction task are therefore the layout mask, the frequency, and the circuit context. Dual-metal-layer pixelated matching networks have been used in fabricated CMOS PAs in GlobalFoundries 22nm FDX+ technology \citep{chu2025deepmatching,chu2025template}. In particular, \citet{chu2025template} use these networks for input, interstage, and output matching in a measured 12--16~GHz PA.

The prediction targets include both EM response and design performance. The EM response is represented by frequency-dependent complex S-parameters, which describe how signals scatter between ports. At each frequency, the full S-matrix has shape $8\times8$. Reciprocity gives $S_{ij}=S_{ji}$; together with layout symmetry, it reduces the matrix to 20 distinct complex entries. The design performance consists of the complex input impedance $Z_{\mathrm{in}}$, together with the real-valued passive efficiency and max gain.

The traditional workflow evaluates a passive matching network in two stages. First, a finite-difference-based EM simulator is run on the layout over frequency to obtain its EM response. Second, this EM response is embedded into the surrounding circuit, and a downstream solver evaluates the resulting design performance under the circuit context. This workflow is physically grounded and widely used in practice, but the EM simulator is computationally expensive and can require up to 20 minutes for a single full-frequency evaluation. As a result, iterative design and large-scale exploration remain slow. Neural modeling of RFIC passives replaces this expensive pipeline by predicting EM response and design performance directly with neural networks, enabling faster evaluation and design iteration.

\section{Method}
\label{sec:method}

\begin{figure*}[t]
    \centering
    \includegraphics[width=0.95\textwidth]{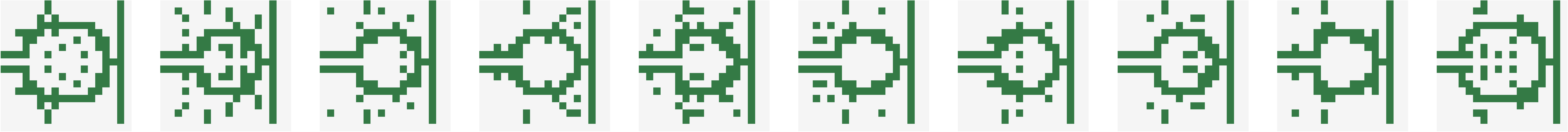}\\
    {\small (a) Contour-seeded, QA metal layer}

    \vspace{2mm}

    \includegraphics[width=0.95\textwidth]{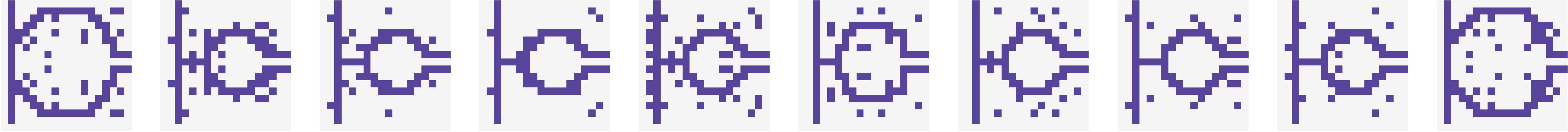}\\
    {\small (b) Contour-seeded, QB metal layer}

    \vspace{2mm}

    \includegraphics[width=0.95\textwidth]{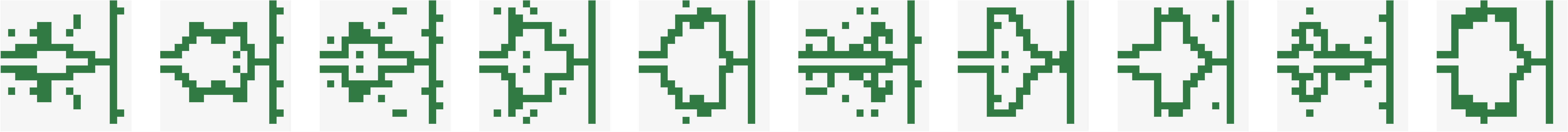}\\
    {\small (c) RandPath-seeded, QA metal layer}

    \vspace{2mm}

    \includegraphics[width=0.95\textwidth]{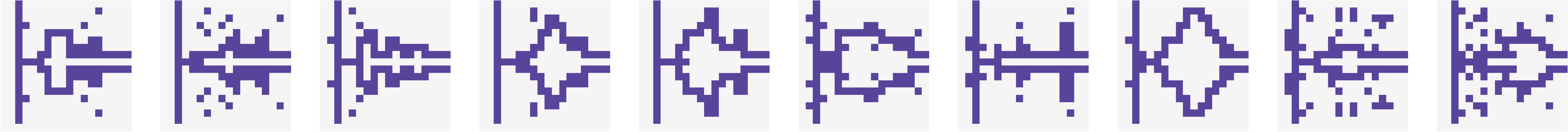}\\
    {\small (d) RandPath-seeded, QB metal layer}
    \caption{Example layouts from the two passive layout distributions.}
    \label{fig:mask_visualization}
\end{figure*}

\paragraph{Problem Formulation.}
We study one passive matching network at a time and train separate models for input, interstage, and output matching. Each passive layout is represented by a binary mask $\mM \in \{0,1\}^{2\times18\times18}$. The masks are drawn from two data distributions, illustrated in \cref{fig:mask_visualization}. Contour-seeded masks start from high-performing expert-designed contours and then add random pixels. RandPath-seeded masks start from random paths between the two ports, mirror the paths to satisfy symmetry, and then add random pixels. These two families cover both structured layouts near established designs and more random connected layouts.

Here, \emph{free-form} denotes pixel-based geometry within the fixed grid and stated structural constraints, allowing layouts beyond predefined geometric templates. This describes the flexibility of the geometric representation, while prediction accuracy on unseen layout distributions is evaluated separately.

For each layout, we sample 200 circuit contexts $\vc=(C_1,C_2)$ by drawing $C_1$ and $C_2$ uniformly at random from $[0,400]$ fF and $[10,400]$ fF, respectively, and then quantizing them to 0.01 fF. Each context is evaluated on the frequency grid $f \in \gF = \{10 + 0.1k \;\text{GHz} : k=0,\ldots,500\}$.
The prediction task is to learn a mapping from $(\mM, f, \vc)$ to the EM response $S(\mM, f) \in \R^{40}$ and the design performance $Y(\mM, f, \vc) \in \R^4$, as defined in \cref{sec:background}.

\begin{figure*}[t]
    \centering
    \includegraphics[width=0.91\textwidth]{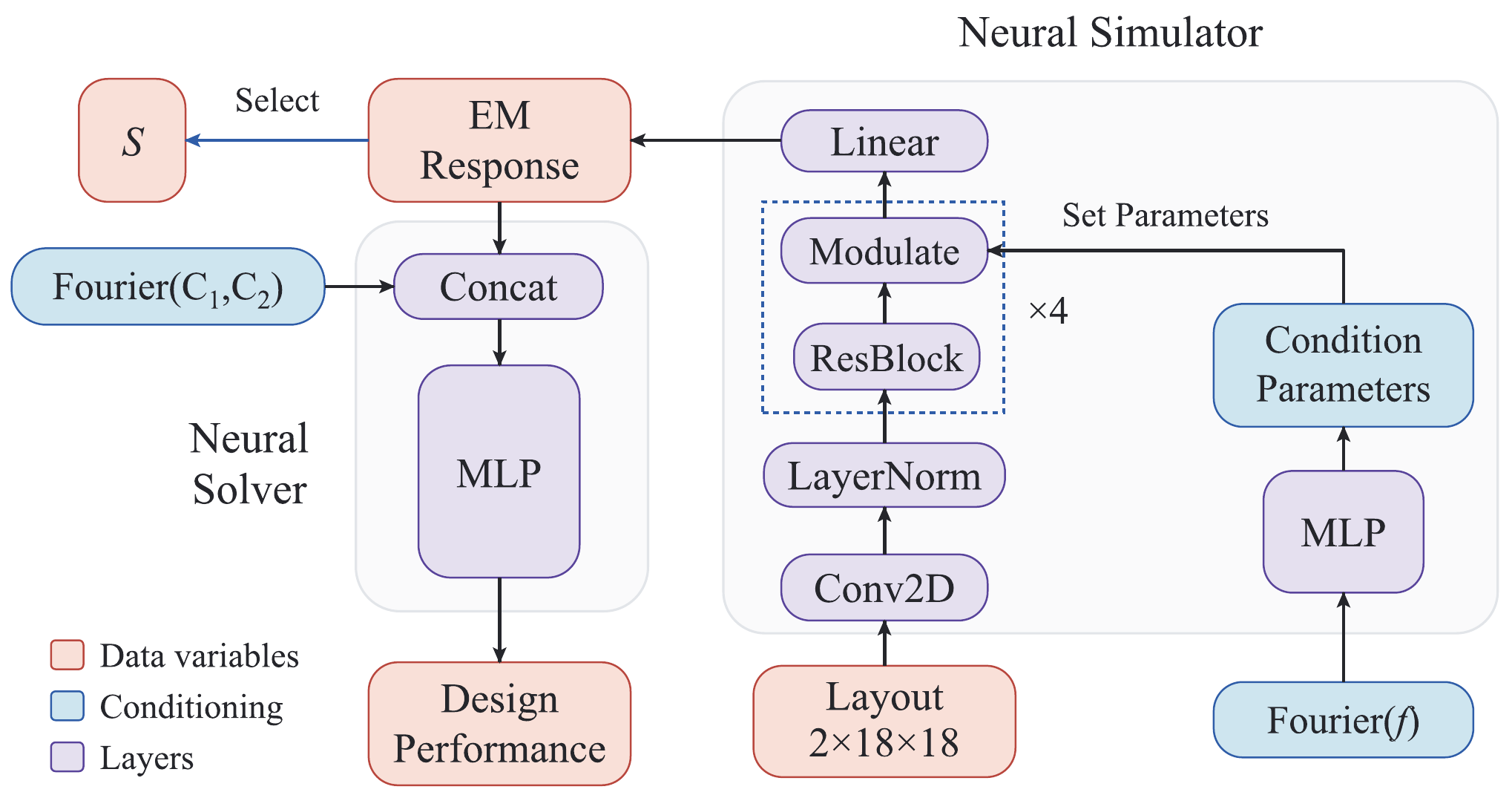}
    \caption{Overview of the proposed model. The frequency-conditioned neural simulator maps the layout $\mM$ and frequency to a neural EM response, consisting of an explicit EM response and a complementary response. The circuit-conditioned neural solver then combines this neural EM response with the circuit context to predict design performance. Frequency information is injected through modulation across four residual blocks.}
    \label{fig:forward_model_arch}
\end{figure*}

\paragraph{Overview.}
Our framework combines a frequency-conditioned \emph{neural simulator} with a circuit-conditioned \emph{neural solver}, as shown in \cref{fig:forward_model_arch}. The neural simulator maps the passive layout $\mM$ and frequency $f$ to a \emph{neural EM response}. This response has two parts: an explicit EM response $\hat S$, supervised against the target S-parameters, and a complementary response $Z_{\mathrm{comp}}$, which preserves additional information useful for downstream design prediction. The neural solver then combines $(\hat S,Z_{\mathrm{comp}})$ with the circuit context $\vc=(C_1,C_2)$ to predict design performance.

\paragraph{Frequency-Conditioned Neural Simulator.}
The neural simulator processes $\mM$ with a convolutional backbone comprising an initial 2D convolution, layer normalization, and four residual blocks \citep{he2016deep}. To condition this shared spatial backbone on frequency, we first compute the Fourier encoding
\begin{equation}
    \phi(f) = [f, \sin(2^0\pi f), \cos(2^0\pi f), \ldots, \sin(2^{K-1}\pi f), \cos(2^{K-1}\pi f)],
\end{equation}
where $K=10$ is the number of Fourier scales. This fixed-size encoding captures both low- and high-frequency variation. Inspired by the affine conditioning in DiT \citep{peebles2023scalable}, we inject frequency information through a modulation layer that follows the feature-wise linear modulation (FiLM) principle \citep{perez2018film}. Conventional CNN FiLM shares each channel's scale and shift across spatial positions. Our layer instead predicts a separate scale and shift for every channel and spatial position from the Fourier-encoded frequency. This design keeps the layout-processing backbone shared across frequencies while allowing intermediate activations to vary smoothly with $f$.
Specifically, for each modulation layer, a conditioning MLP maps $\phi(f)$ to modulation parameters $\bm{\gamma}(f)$ and $\bm{\beta}(f)$. Let $\vh(\mM) \in \R^{d}$ denote the vectorized input feature tensor to one such layer. Then $\bm{\gamma}(f), \bm{\beta}(f) \in \R^{d}$, and the modulated feature is
\begin{equation}
    \widetilde{\vh}(\mM,f) = \bm{\gamma}(f) \odot \vh(\mM) + \bm{\beta}(f),
\end{equation}
where $\odot$ denotes elementwise multiplication. Continuous frequency conditioning allows queries between the discrete training frequencies, as evaluated within the training band in \cref{sec:experiments}. Since the conditioning parameters depend only on frequency, they can be cached when evaluating multiple layouts at the same frequency.

\paragraph{Circuit-Conditioned Neural Solver.}
The neural solver predicts design performance from the neural EM response and circuit context. We encode $C_1$ and $C_2$ with the same Fourier feature construction, using $K=5$ for each scalar, concatenate the resulting embedding with the neural EM response at each frequency, and process the concatenated representation with an MLP to obtain $\hat Y$. This design mirrors the traditional workflow in \cref{sec:background}: the simulator provides a learned surrogate for EM analysis, and the solver maps the resulting neural EM response into downstream design quantities under the specified circuit context.

The learned solver also improves robustness to imperfect EM predictions. As shown in \cref{fig:solver_instability}, a ground-truth downstream solver is highly sensitive to perturbations in its EM-response input. Therefore, high EM-response accuracy alone is insufficient for accurate design-performance prediction when the traditional solver is applied. Similar observations have also been reported in different settings by \citet{carlucci2023load,carlucci2025robustness}. The learned neural solver can adapt to errors in the neural EM response, enabling  design performance prediction even with imperfect EM response.

\paragraph{Joint Training Objective.}
We train the simulator and solver end-to-end with normalized mean squared error losses. Let $\hat S$ and $S$ denote the predicted and ground-truth EM responses, and let $\hat Y$ and $Y$ denote the predicted and ground-truth design performances. We define
\begin{align}
    \Ls_{\mathrm{EM}} &= \frac{1}{|\gF|} \sum_{f \in \gF} \lVert \hat S(f) - S(f) \rVert_2^2, \\
    \Ls_{\mathrm{design}} &= \frac{1}{|\gC|\,|\gF|} \sum_{\vc \in \gC} \sum_{f \in \gF} \lVert \hat Y(\vc,f) - Y(\vc,f) \rVert_2^2,
\end{align}
where $\gF$ is the frequency grid and $\gC$ is the sampled set of circuit contexts. The prediction loss is
\begin{equation}
    \Ls_{\mathrm{pred}} = \Ls_{\mathrm{EM}} + \Ls_{\mathrm{design}}.
\end{equation}
This joint supervision keeps the simulator faithful to the EM target while encouraging representations that support accurate downstream design-performance prediction.

\paragraph{Dependence Regularization.}
The complementary response $Z_{\mathrm{comp}}$ is intended to capture information beyond the explicit EM-response, thus reducing the information bottleneck due to the imperfect EM response. To encourage independence between $Z_{\mathrm{comp}}$ and $\hat{S}$, we regularize $Z_{\mathrm{comp}}$ additionally to reduce its correlation with $\hat{S}$. Let $\vu$ denote the simulator feature before the final linear projections, and let
\begin{equation}
    \hat S = \mW_S \vu + \vb_S, \qquad Z_{\mathrm{comp}} = \mW_Z \vu + \vb_Z.
\end{equation}
Since both are affine transformations of the same feature, their covariance has the closed form
\begin{equation}
    \Cov(\hat S, Z_{\mathrm{comp}}) = \mW_S \, \Cov(\vu) \, \mW_Z^\top.
\end{equation}
For a minibatch with feature vectors $\vu_1, \ldots, \vu_n$, we estimate $\Cov(\vu)$ by the empirical covariance
\begin{equation}
    \widehat{\Cov}(\vu) = \frac{1}{n-1} \sum_{i=1}^{n} (\vu_i - \bar{\vu})(\vu_i - \bar{\vu})^\top,
    \qquad \bar{\vu} = \frac{1}{n} \sum_{i=1}^{n} \vu_i.
\end{equation}
We then define
\begin{equation}
    \Ls_{\mathrm{corr}} = \frac{1}{d_S d_Z} \left\lVert \operatorname{sg}(\mW_S) \, \operatorname{sg}(\widehat{\Cov}(\vu)) \, \mW_Z^\top \right\rVert_F^2,
\end{equation}
where $d_S$ and $d_Z$ are the dimensions of $\hat{S}$ and $Z_{\mathrm{comp}}$, respectively, $\lVert \cdot \rVert_F$ denotes the Frobenius norm, and $\operatorname{sg}(\cdot)$ denotes stop gradient. This regularization thus only affects the complementary response and put no constraint on the prediction of the explicit EM response $\hat{S}$. The full training objective is
\begin{equation}
    \Ls = \Ls_{\mathrm{pred}} + \reg \, \Ls_{\mathrm{corr}}.
\end{equation}
As shown in \cref{sec:experiments}, this regularization substantially reduces interdependence with little loss in accuracy.

\section{Experiments}
\label{sec:experiments}

\paragraph{Experimental Details.}
We split the full dataset of 61{,}515 passive layouts into training, validation, and test partitions in a $7{:}1{:}2$ ratio.
During training, all target variables are normalized by subtracting the dataset mean and dividing by the dataset standard deviation. Conditioning variables are normalized to $[-1,1]$ according to their ranges.
We also apply layout augmentation with random shifts, flips, and right-angle rotations. We use $R^2$ as the prediction accuracy metric, defined as $R^2 = 1 - \frac{\sum_i (y_i - \hat y_i)^2}{\sum_i (y_i - \bar y)^2}$, where $y_i$ and $\hat y_i$ are the ground-truth and predicted values, respectively, and $\bar y$ is the mean of the ground-truth values. We report $R^2$ for both EM response and design performance targets, denoted as $R^2_{\mathrm{EM}}$ and $R^2_{\mathrm{design}}$, respectively. When the target variable has multiple dimensions, e.g., $S \in \R^{40}$, we compute $R^2$ for each dimension and report the average.

We study two model scales: a base model with 5.4M parameters and a large model with 21M parameters.
Both are optimized with AdamW, using learning rate $10^{-4}$, weight decay $10^{-2}$, and a cosine annealing schedule that decays the learning rate to $10^{-6}$ over training.
For efficient storage and loading, we store the dataset in Zarr format. The stored dataset occupies about 200 GB after compressing 32-bit floating-point values to 16 bits, followed by Bitshuffle~\citep{masui2017bitshuffle} and Blosc compression.
Training uses distributed data parallelism over 8 NVIDIA GeForce RTX 2080 Ti GPUs with batch size 4 per GPU, and we report test results from the checkpoint with the lowest validation loss.
Unless stated otherwise, all models are trained for 500 epochs, with validation after every epoch. This takes roughly 45 hours for the 5.4M model and roughly 120 hours for the 21M model to train end-to-end.
After initial tuning on the 5.4M model with few epochs, we keep the hyperparameters fixed across all model scales and matching settings, without further tuning. We did not perform cherry-picking of random seeds, and all results are from a single end-to-end run per model, due to the high computational cost. Nevertheless, the results are stable across model scales and matching settings in independent runs.

\paragraph{Main Results Across Model Scales and Matching Settings.}
\cref{tab:main_results} compares the 5.4M and 21M models across input, interstage, and output matching. Results are reported on the full distribution and on the two constituent distributions.
Scaling from 5.4M to 21M consistently improves both EM response and design performance prediction across all three matching settings.
These gains suggest that model capacity is an important factor for free-form RFIC passive modeling. Further scaling may yield additional improvements; we leave it for future work due to the high cost.
\Cref{app:backbone_comparison} further compares parameter-matched plain CNN and residual backbones trained for 100 epochs on input matching. Their comparable aggregate accuracy indicates that residual blocks are not essential to high performance in this setting.

\begin{table*}[t]
\centering
\caption{Main results for the 5.4M and 21M models across input, interstage, and output matching. We report EM response $R^2_{\mathrm{EM}}$ and design performance $R^2_{\mathrm{design}}$ on the full distribution and on its two constituent distributions. Scaling to 21M improves both metrics across all three matching settings.}
\label{tab:main_results}
\begin{adjustbox}{max width=\textwidth,center}
\small
\setlength{\tabcolsep}{5pt}
\begin{tabular}{cccccccc}
\toprule
\multirow{2}{*}{\#param} & \multirow{2}{*}{Matching} & \multicolumn{2}{c}{Full} & \multicolumn{2}{c}{Contour-seeded} & \multicolumn{2}{c}{RandPath-seeded} \\
\cmidrule(lr){3-4}\cmidrule(lr){5-6}\cmidrule(lr){7-8}
 &  & $R^2_{\mathrm{EM}}$ & $R^2_{\mathrm{design}}$ & $R^2_{\mathrm{EM}}$ & $R^2_{\mathrm{design}}$ & $R^2_{\mathrm{EM}}$ & $R^2_{\mathrm{design}}$ \\
\midrule
5.4M  & \multirow{2}{*}{input}      & 0.9879 & 0.9439 & 0.9885 & 0.9327 & 0.9940 & 0.9442 \\
21M &       & \textbf{0.9944} & \textbf{0.9666} & \textbf{0.9934} & \textbf{0.9782} & \textbf{0.9968} & \textbf{0.9641} \\
\midrule
5.4M  & \multirow{2}{*}{interstage} & 0.9844 & 0.9282 & 0.9864 & 0.8759 & 0.9920 & 0.9191 \\
21M &  & \textbf{0.9932} & \textbf{0.9394} & \textbf{0.9921} & \textbf{0.9647} & \textbf{0.9962} & \textbf{0.9381} \\
\midrule
5.4M  & \multirow{2}{*}{output}     & 0.9878 & 0.9503 & 0.9881 & 0.9350 & 0.9937 & 0.9492 \\
21M &      & \textbf{0.9939} & \textbf{0.9714} & \textbf{0.9936} & \textbf{0.9810} & \textbf{0.9968} & \textbf{0.9723} \\
\bottomrule
\end{tabular}
\end{adjustbox}
\end{table*}

\begin{figure*}[t]
    \centering
    \includegraphics[width=0.95\textwidth]{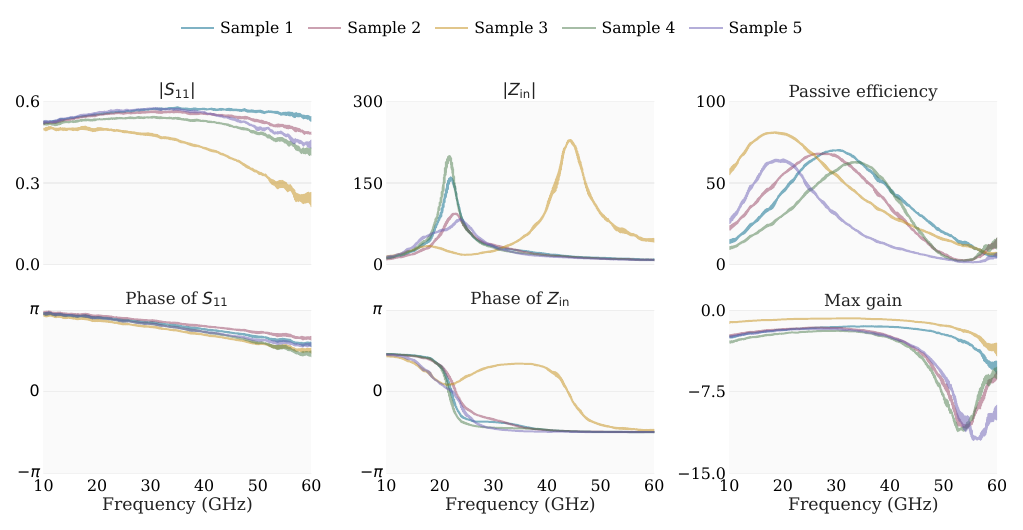}
    \caption{Predictions of the 21M input matching model on a frequency grid with 0.001~GHz spacing. Wider curve means the prediction changes less smoothly.}
    \label{fig:prediction_over_frequency_xl_input_matching_samples}
\end{figure*}

\paragraph{Interpolation Between Training Frequencies.}
\cref{fig:prediction_over_frequency_xl_input_matching_samples} evaluates the 21M input matching model on a frequency grid that is $100\times$ denser than the training grid, using a query step of 0.001~GHz, for five randomly sampled test layouts. Corresponding interstage and output matching results are provided in \cref{app:frequency_dense}.
This evaluation queries unseen frequencies within the 10--60~GHz training band and assesses interpolation. We do not evaluate extrapolation because RFIC usually has an operating range and does not need to extrapolate to other bands.
The model remains well behaved under this much finer frequency query, indicating that the frequency-conditioned simulator interpolates smoothly between training frequencies.
The denser evaluation also reveals visibly less smooth responses in high frequencies, possibly due to the increased sensitivity of the targets.
This observation suggests that future work could explore non-uniform frequency sampling during training, with denser coverage in high-frequency regions.

\begin{figure*}[t]
    \centering
    \includegraphics[width=0.95\textwidth]{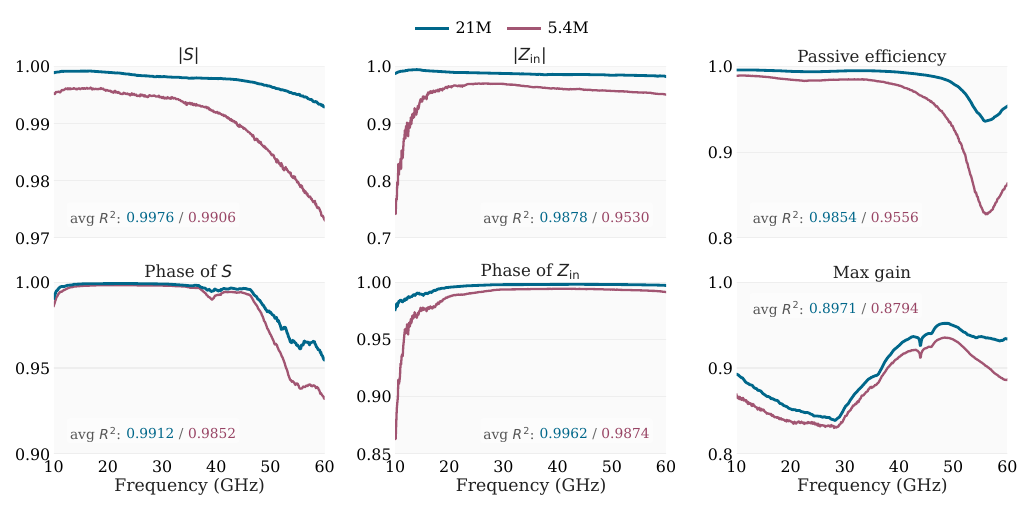}
    \caption{Per-target $R^2$ over frequency for the 21M and 5.4M input matching models.}
    \label{fig:r2_over_frequency_ours_21m_input_matching_full}
\end{figure*}

\paragraph{Per-Target Accuracy Across Frequency.}
\cref{fig:r2_over_frequency_ours_21m_input_matching_full} compares the 21M and 5.4M input matching models on the full distribution, broken down by target and frequency. Corresponding interstage and output matching results are provided in \cref{app:frequency_scale}.
Scaling to 21M improves all targets across frequency band, with the clearest gains in more difficult prediction regimes where the 5.4M model struggles, including harder targets and frequency ranges. Among the six targets, max gain remains the most difficult for the 21M model, with accuracy about 10\% lower than the other targets.

\begin{figure*}[t]
    \centering
    \includegraphics[width=0.95\textwidth]{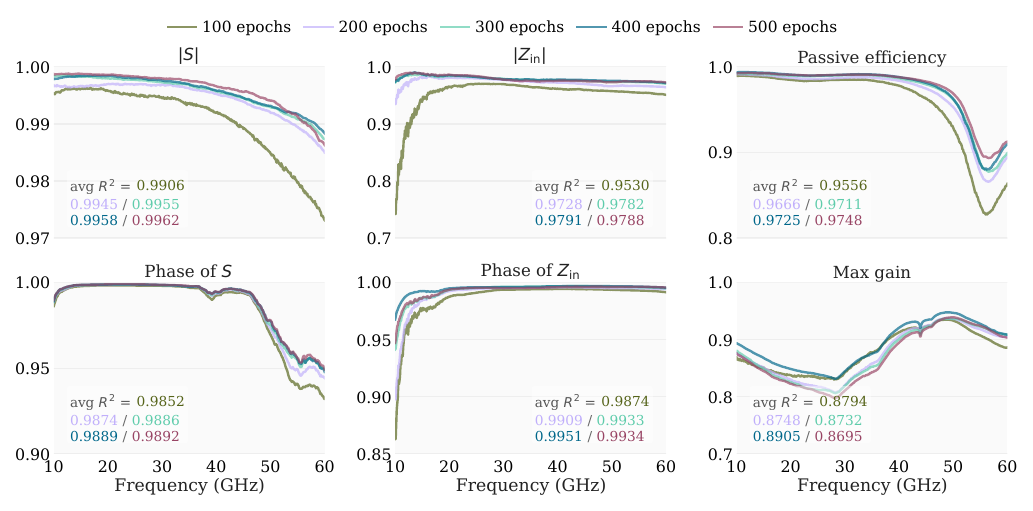}
    \caption{Per-target $R^2$ over frequency for the 5.4M input matching model across training epochs.}
    \label{fig:base_continuous_r2_over_frequency_input_matching}
\end{figure*}

\paragraph{Effect of Training Compute Scaling.}
\cref{fig:base_continuous_r2_over_frequency_input_matching} tracks the 5.4M input matching model across training epochs. Corresponding interstage and output matching results are provided in \cref{app:frequency_training}.
Longer training improves accuracy across all targets, with the largest gains on challenging targets and frequency ranges. These results demonstrate the benefit of increasing training compute.

\begin{figure}[t]
    \centering
    \includegraphics[width=0.95\linewidth]{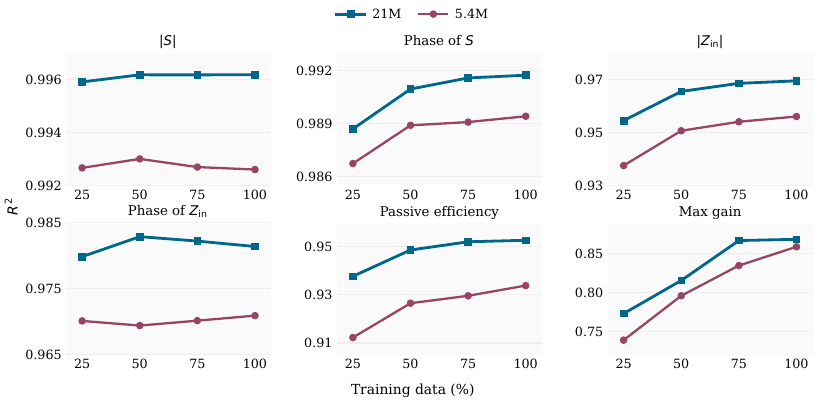}
    \caption{In-distribution data scaling for input matching. Both models are trained and tested on RandPath-seeded layouts with a fixed training budget of 52{,}450 optimizer steps. Each panel reports test $R^2$ for one target. Panel-specific $y$ limits differ.}
    \label{fig:randpath_data_scaling}
\end{figure}

\paragraph{Effect of Dataset Scaling.}
We train both model sizes on nested subsets containing 25\%, 50\%, 75\%, and 100\% of the RandPath-seeded training layouts for input matching, using a fixed number of steps equivalent to 50 full-data epochs.
Dataset scaling consistently improves $R^2$ for almost all targets (\cref{fig:randpath_data_scaling}). As with training compute scaling, the gains are most pronounced for challenging targets with lower $R^2$.

\begin{figure*}[t]
    \centering
    \includegraphics[width=0.95\textwidth]{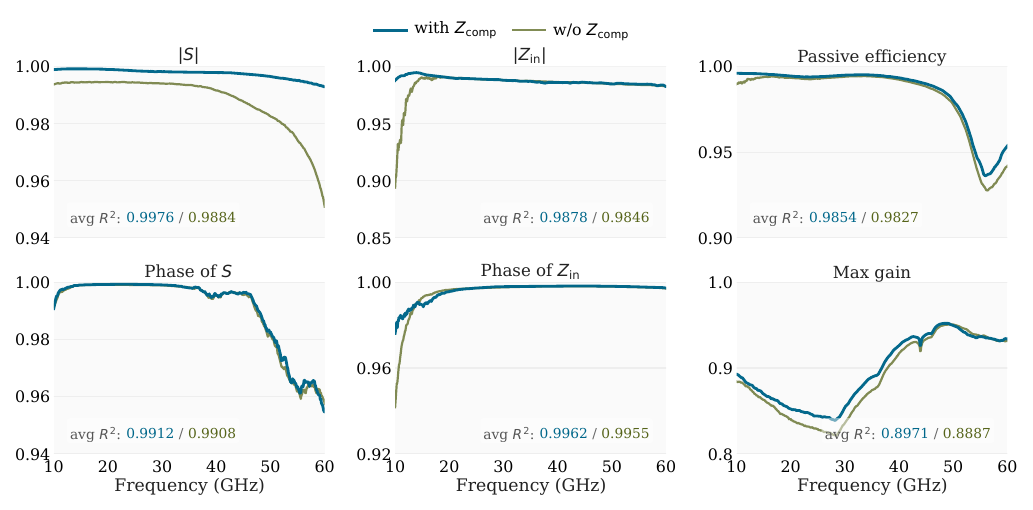}
    \caption{Effect of the complementary response for the 21M input matching model.}
    \label{fig:r2_over_frequency_zcomp_ablation_input_matching_full}
\end{figure*}

\paragraph{Effect of the Complementary Response.}
\cref{fig:r2_over_frequency_zcomp_ablation_input_matching_full} compares the 21M input matching model with and without the complementary response.
Adding $Z_{\mathrm{comp}}$ improves the average $R^2$ for every target.
The largest gain appears on $|S|$, where removing $Z_{\mathrm{comp}}$ causes a pronounced high-frequency drop. This is possibly because that without $Z_{\mathrm{comp}}$, the model has to adapt to the sensitive dependence between the design performance and $S$, thus preventing it from learning better $\hat{S}$.
For $|Z_{\mathrm{in}}|$ and phase of $Z_{\mathrm{in}}$, $Z_{\mathrm{comp}}$ prevents sharp degradation in the low-frequency region. The remaining targets also receive consistent gains from $Z_{\mathrm{comp}}$ across the frequency band.

\begin{figure}[t]
    \centering
    \includegraphics[width=0.72\linewidth]{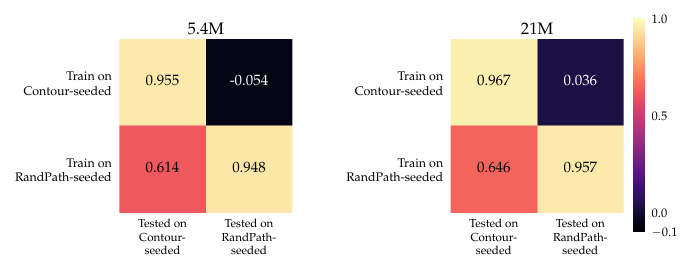}
    \caption{Generalization to unseen mask distributions between the Contour-seeded and RandPath-seeded families. Each cell reports $R^2_{\mathrm{design}}$. Models are trained for 100 epochs for this analysis.}
    \label{fig:cross_distribution_generalization}
\end{figure}

\paragraph{Generalization to Unseen Layout Distributions.}
We test cross-distribution generalization between the two constituent distributions, Contour-seeded and RandPath-seeded, which have different layout generation procedures and layout characteristics.
As shown in \cref{fig:cross_distribution_generalization}, the direction of transfer matters substantially.
Models trained on Contour-seeded masks perform well on the same family ($R^2_{\mathrm{design}}=0.955$ and $0.967$ for 5.4M and 21M), but generalize poorly to the more random RandPath-seeded family ($R^2_{\mathrm{design}}=-0.054$ and $0.036$).
By contrast, models trained on RandPath-seeded masks transfer much better to the unseen Contour-seeded family ($R^2_{\mathrm{design}}=0.614$ and $0.646$).
Thus, training on the more random mask distribution provides stronger out-of-distribution generalization, although it still does not match same-family performance. This suggests that broader and more randomized layout distributions may improve generalization to unseen layouts compared with training only on established layout families.

\begin{figure}[t]
    \centering
    \begin{minipage}[t]{0.49\linewidth}
        \centering
        \includegraphics[width=.9\linewidth]{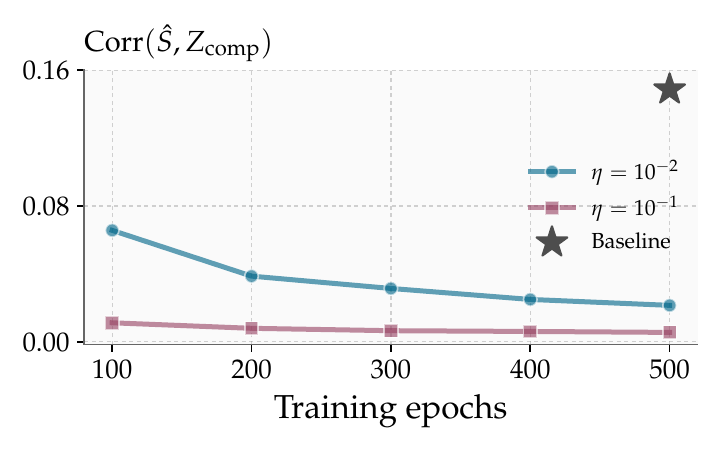}
    \end{minipage}
    \hfill
    \begin{minipage}[t]{0.49\linewidth}
        \centering
        \includegraphics[width=.9\linewidth]{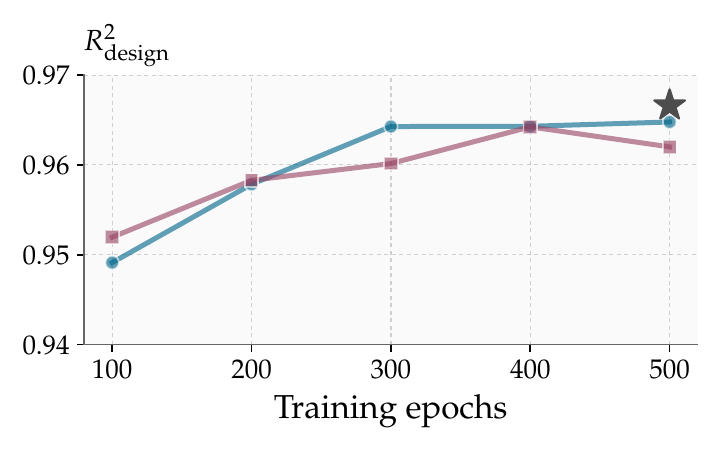}
    \end{minipage}
    \caption{Effect of dependence regularization. $\mathrm{Corr}(\hat S, Z_{\mathrm{comp}})$ is the average correlation.}
    \label{fig:paramnet_epoch_curves}
\end{figure}

\paragraph{Effect of Dependence Regularization.}
\cref{fig:paramnet_epoch_curves} evaluates the effect of the dependence regularization.
With regularization ($\lambda_{\mathrm{corr}}=0.1$), $\mathrm{Corr}(\hat S, Z_{\mathrm{comp}})$ decreases from $0.149$ to $0.006$, corresponding to a 96\% reduction.
This reduction comes with only a small change in design performance, with $R^2_{\mathrm{design}}$ moving from $0.967$ to $0.964$.
These results show that the regularizer yields a cleaner decomposition of the neural EM response while preserving design-performance accuracy.
A per-target accuracy breakdown across regularization strengths is provided in \cref{app:covreg_frequency}.

\section{Discussion}
\label{sec:discussion}

\paragraph{Limitations.}
Our study focuses on one RFIC passive setting with a fixed $18\times18$ layout grid, two metal layers, and three PA matching placements. The models are trained separately for input, interstage, and output matching, and the data cover only two layout-generation families. Cross-family accuracy remains below in-distribution accuracy, and the current evidence does not establish generalization to broader layout distributions. The current checkpoints are not expected to transfer to different layout resolutions or process technologies without adaptation or retraining. The dense-frequency evaluation demonstrates interpolation within the training band rather than extrapolation beyond it. Training is expensive, particularly for the 21M model, and we do not quantify variability across random seeds.
The predicted EM responses do not guarantee physical properties such as passivity; \Cref{app:passivity} quantifies violations of passivity across all three matching settings and shows lower violation rates with better models.

\paragraph{Future Work.}
\textit{(a) Flexible layout sizes.} Even though this work removes the restriction from a templated layout, the fixed layout size still restricts the design space. Encoding layouts as variable-length sequences of patch tokens \citep{dehghani2023navit} could further relax this constraint and support an even broader design space.
\textit{(b) Further scaling.} We demonstrate that larger models (\cref{tab:main_results}), longer training (\cref{fig:base_continuous_r2_over_frequency_input_matching}), and more training data (\cref{fig:randpath_data_scaling}) improve accuracy consistently. These gains motivate further scaling neural simulation models in these directions.
\textit{(c) Broader conditioning.} This work provides a general way to condition neural simulation models; extending the current conditioning to other properties, \eg material effects, could potentially allow one simulator to cover more diverse settings.
\textit{(d) Randomized training data.} Stronger transfer from RandPath-seeded than Contour-seeded training (\cref{fig:cross_distribution_generalization}) suggests that broader training-distribution coverage should accompany increases in dataset size. For broad free-form generalization, this motivates prioritizing randomized simulation data over exclusive reliance on template-seeded layouts \citep{chu2025deepmatching,chu2025template} or predefined transformer families \citep{lee2022neural,he2025motif}.

\section{Conclusion}
\label{sec:conclusion}

We present an end-to-end neural modeling framework for free-form RFIC passive matching networks. The framework combines a frequency-conditioned neural simulator with a circuit-conditioned neural solver, enabling direct prediction of both EM response and design performance under varying frequency and circuit contexts. On a large-scale dataset of free-form RFIC passives, the proposed 21M model achieves high accuracy across input, interstage, and output matching settings, while providing orders-of-magnitude faster evaluation than traditional EM simulation. These results suggest that neural EM and design-performance modeling can substantially accelerate RFIC passive design-space exploration, with future extensions toward more diverse layouts, flexible layout dimensions, and broader circuit settings.

\clearpage
\message{^^JLASTBODYPAGE \number\numexpr\value{page}-1\relax^^J}

\section*{LLM Disclosure}

This paper was assisted by PIRA~\citep{pira}, a research-assistant system powered by GPT-6-Astra and GPT-6.1-Sol. The assistance included implementation aids and writing polishes. The authors are fully responsible for the presented final content.

\bibliography{references}
\bibliographystyle{IEEEtranN}

\message{^^JLASTREFERENCESPAGE \thepage^^J}

\ifbool{includeappendix}{%
	\clearpage
	\appendix
	\begingroup
\raggedbottom
\setlength{\intextsep}{4pt}

\crefname{appendix}{Appendix}{Appendices}
\Crefname{appendix}{Appendix}{Appendices}
\crefalias{section}{appendix}
\crefalias{subsection}{appendix}

\section{Related Work}
\label{sec:related_work}

\paragraph{Neural Modeling of RFIC Passives and Downstream Applications.}
The high cost of electromagnetic (EM) simulation has motivated neural surrogates for RF and microwave passive components.
Existing methods learn geometry- and frequency-conditioned mappings to EM responses, including multiphysics passive models \citep{yan2018multi}, RF-device S-parameter models \citep{guan2021modeling}, on-chip inductor models with dummy metal fillings \citep{li2022deep}, recurrent surrogates with domain confinement for passive microwave circuits \citep{sahu2025surrogate}, and uncertainty-aware Bayesian online learning for RF passive components \citep{zhang2026efficient}.
These models primarily accelerate EM-response evaluation, thereby supporting design-space exploration, optimization, uncertainty analysis, and inverse synthesis.
For free-form and multi-port structures, \citet{karahan2024deep} use a deep forward model of scattering parameters to support inverse design of RF/sub-THz passives and integrated circuits.
Our framework differs from this line of work by jointly predicting the explicit EM response and the downstream design performance of free-form PA matching networks under varying frequency and circuit contexts.

\paragraph{RFIC Design with Neural Methods.}
Neural methods have also been used directly for RFIC design and synthesis.
Early direct-synthesis methods map target specifications to transformer geometries for RF/mm-wave matching networks \citep{munzer2020residual,er2021deep}, and \citet{lee2022neural} propose an automated transformer matching-circuit synthesizer for RF amplifier design.
Subsequent tool-oriented and data-efficient methods include PACOSYT for inductor and transformer synthesis \citep{passos2023pacosyt}, transfer learning for fast migration of transformer matching networks across technology nodes \citep{chu2024transfer}, and multi-template transformer synthesis with frequency-domain self-transfer learning \citep{he2025motif}.
For broader free-form and circuit-co-designed synthesis, learned forward models have enabled inverse design of mm-wave passives and PAs \citep{liu2022deepinverse,karahan2024deep}, physics-augmented modeling and synthesis in PulseRF \citep{chae2024pulserf}, fully AI-generated passive networks in a D-band PA \citep{chai2025dband}, diffusion-based target-to-EM-structure generation \citep{guo2025dallem}, and end-to-end RF/mm-wave automation flows such as COmPOSER \citep{ghosh2026composer}.
Closest to our PA matching-network setting, recent 22nm FDX+ demonstrations combine template-guided or template-seeded pixelated dual- and multi-metal-layer passive networks with CNN-based EM emulators to realize measured 6G FR3 CMOS PAs \citep{chu2025deepmatching,chu2025template}.
These approaches show that neural models can move beyond fast evaluation toward automated RFIC design, but they mainly optimize or synthesize structures through EM-response targets, inverse-design objectives, or full design-flow validation.

\section{Additional Frequency-Dependent Analyses}
\label{app:appendix}
\label{app:frequency_analyses}

This appendix complements the input-matching analyses in \cref{fig:r2_over_frequency_ours_21m_input_matching_full,fig:base_continuous_r2_over_frequency_input_matching,fig:prediction_over_frequency_xl_input_matching_samples,fig:r2_over_frequency_zcomp_ablation_input_matching_full,fig:paramnet_epoch_curves} with corresponding frequency-dependent analyses for the interstage and output matching settings.
We group the figures by objective: scaling behavior across frequency (\cref{app:frequency_scale}), training evolution across frequency (\cref{app:frequency_training}), dense-frequency-query visualizations for the 21M model (\cref{app:frequency_dense}), and a frequency-level breakdown of dependence regularization (\cref{app:covreg_frequency}).

\clearpage
\subsection{Frequency-Level Scaling Analysis}
\label{app:frequency_scale}

\vfill
\begin{figure}[H]
    \centering
    \includegraphics[width=0.95\textwidth]{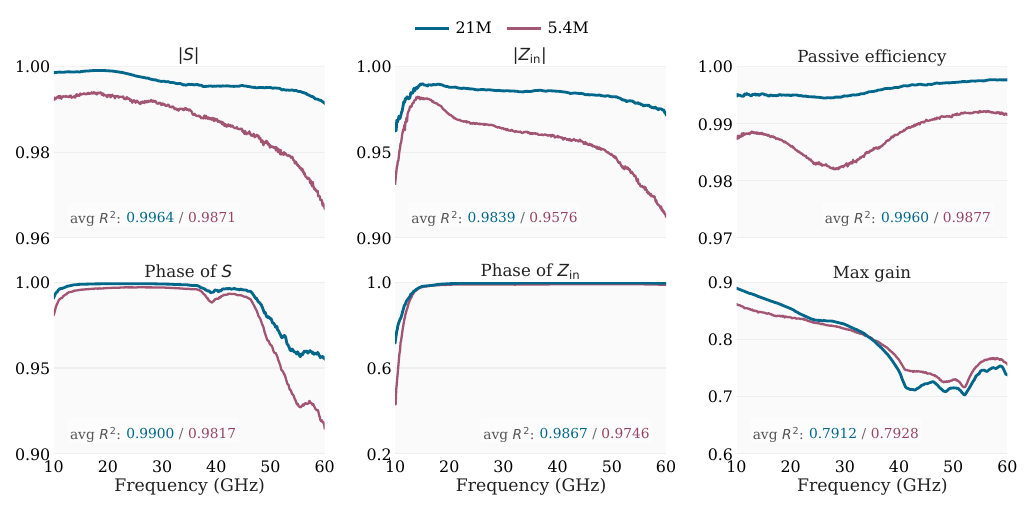}
    \caption{Per-target $R^2$ over frequency for the 21M and 5.4M interstage matching models on the full distribution.}
    \label{fig:r2_over_frequency_ours_21m_interstage_matching_full}
\end{figure}

\vfill
\begin{figure}[H]
    \centering
    \includegraphics[width=0.95\textwidth]{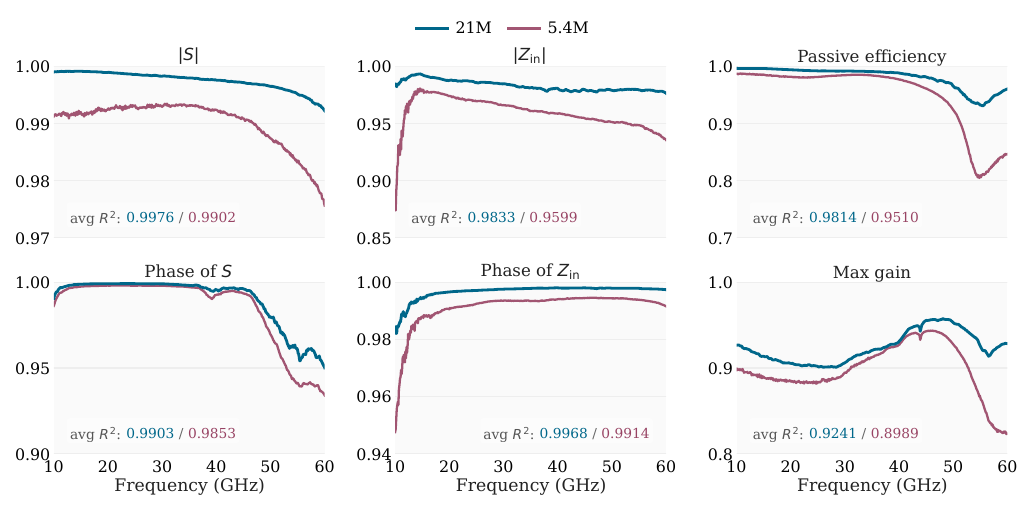}
    \caption{Per-target $R^2$ over frequency for the 21M and 5.4M output matching models on the full distribution.}
    \label{fig:r2_over_frequency_ours_21m_output_matching_full}
\end{figure}

\vfill
\clearpage
\subsection{Frequency-Level Training Analysis}
\label{app:frequency_training}

\vfill
\begin{figure}[H]
    \centering
    \includegraphics[width=0.95\textwidth]{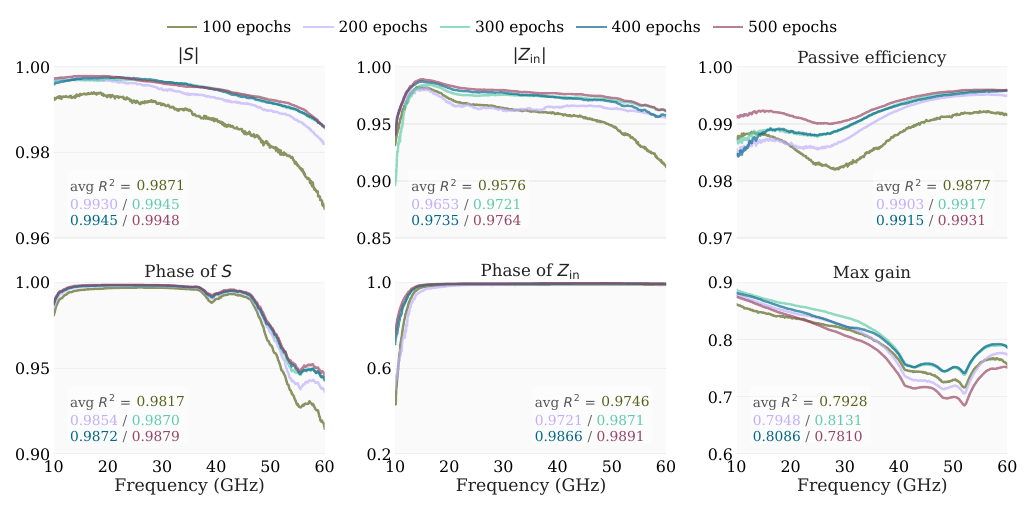}
    \caption{Per-target $R^2$ over frequency for the 5.4M interstage matching model across training epochs on the full distribution.}
    \label{fig:base_continuous_r2_over_frequency_interstage_matching}
\end{figure}

\vfill
\begin{figure}[H]
    \centering
    \includegraphics[width=0.95\textwidth]{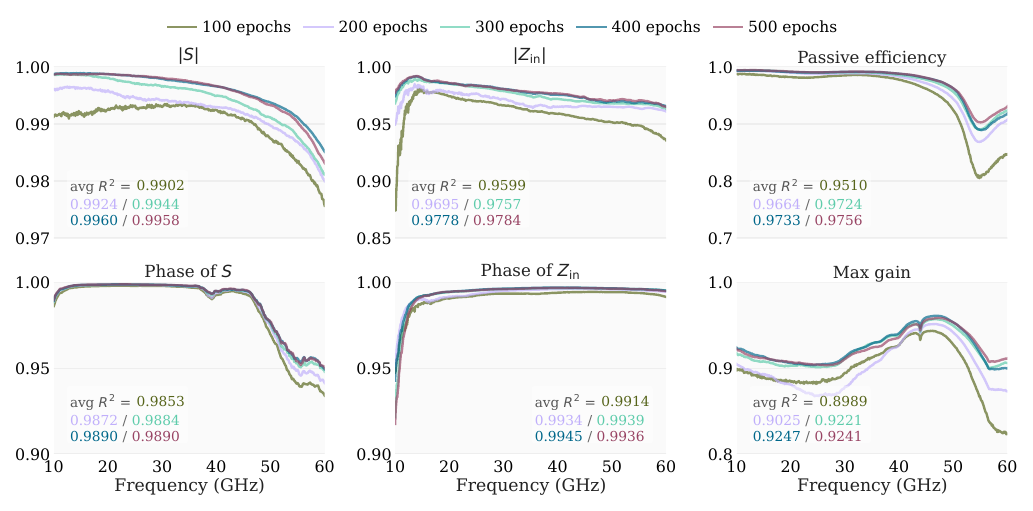}
    \caption{Per-target $R^2$ over frequency for the 5.4M output matching model across training epochs on the full distribution.}
    \label{fig:base_continuous_r2_over_frequency_output_matching}
\end{figure}

\vfill
\clearpage
\subsection{Dense-Frequency Query Analysis}
\label{app:frequency_dense}

\vfill
\begin{figure}[H]
    \centering
    \includegraphics[width=0.95\textwidth]{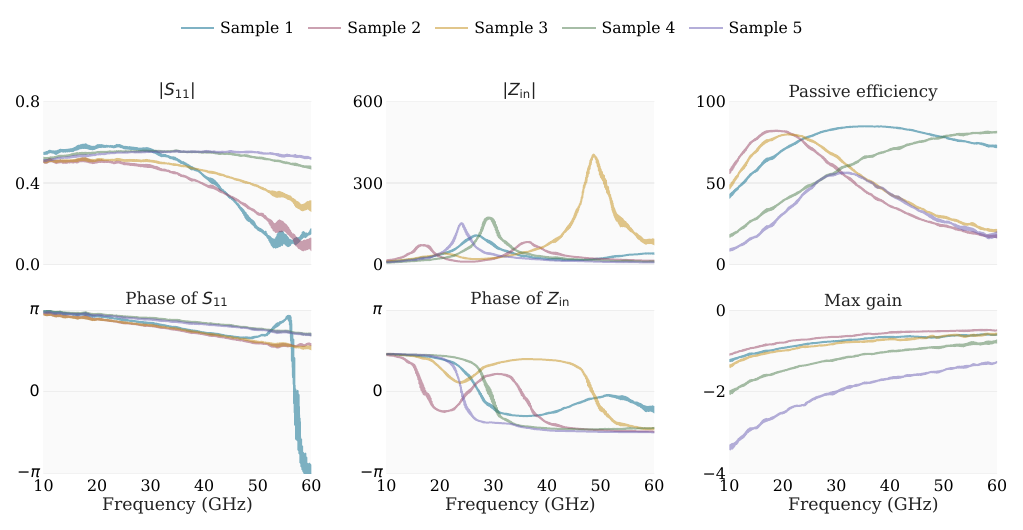}
    \caption{Predictions of the 21M interstage matching model on a frequency grid with 0.001~GHz spacing, which is $100\times$ denser than the training grid.}
    \label{fig:prediction_over_frequency_xl_interstage_matching_samples}
\end{figure}

\vfill
\begin{figure}[H]
    \centering
    \includegraphics[width=0.95\textwidth]{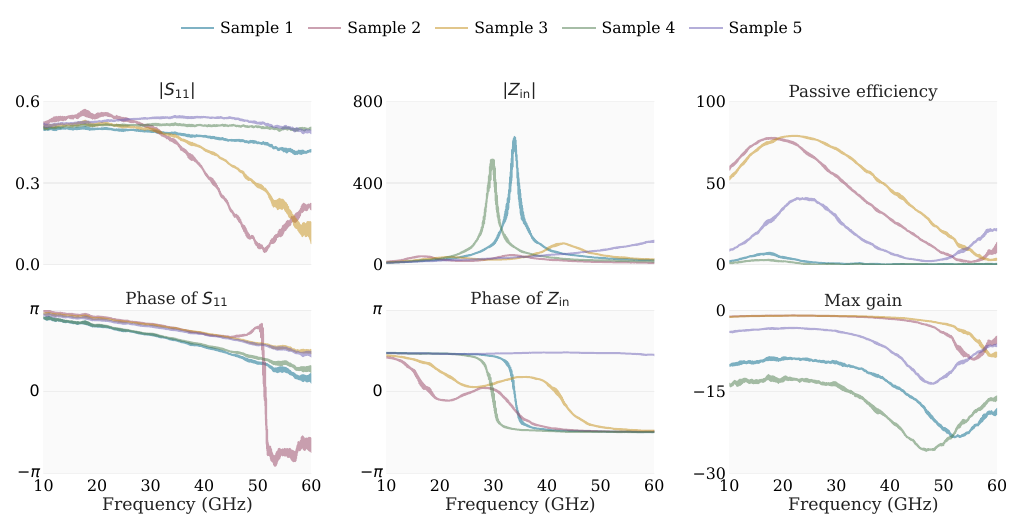}
    \caption{Predictions of the 21M output matching model on a frequency grid with 0.001~GHz spacing, which is $100\times$ denser than the training grid.}
    \label{fig:prediction_over_frequency_xl_output_matching_samples}
\end{figure}

\vfill
\clearpage
\subsection{Dependence-Regularization Frequency Analysis}
\label{app:covreg_frequency}

\begin{figure}[H]
    \centering
    \includegraphics[width=0.95\textwidth]{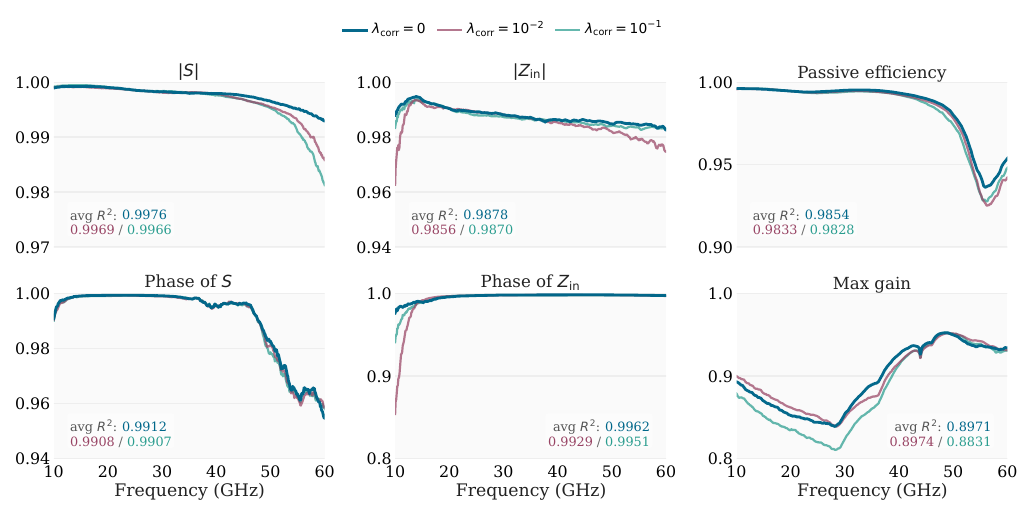}
    \caption{Per-target $R^2$ over frequency for the 21M input matching model under different dependence-regularization strengths on the full distribution.}
    \label{fig:r2_over_frequency_lambda_corr_input_matching_full}
\end{figure}

\section{Sensitivity to the Simulator Backbone}
\label{app:backbone_comparison}

We evaluate whether the residual backbone is essential to the performance of the conditional neural simulator--solver framework.
We replace the neural simulator's residual backbone with a shallow convolutional neural network (CNN) of comparable parameter count, retaining frequency conditioning and the neural solver.
Both models are trained for 100 epochs on the full mixed-distribution training split for input matching, with seed 1, batch size 4 per device, learning rate $10^{-4}$, and weight decay $10^{-2}$.
We evaluate the checkpoint with the lowest validation loss on the full test distribution.

\begin{table}[htbp]
    \centering
    \caption{Simulator-backbone comparison for input matching after 100 training epochs. Both models retain the conditional simulator--solver framework.}
    \label{tab:backbone_comparison}
    \begin{tabular}{lrrr}
        \toprule
        Backbone & Parameters & $R^2_{\mathrm{EM}}$ & $R^2_{\mathrm{design}}$ \\
        \midrule
        Plain CNN & 5,347,884 & 0.98839 & 0.94128 \\
        Residual & 5,383,468 & 0.98792 & 0.94386 \\
        \bottomrule
    \end{tabular}
\end{table}

\Cref{tab:backbone_comparison} shows comparable aggregate accuracy for the two backbones.
The CNN achieves slightly higher $R^2_{\mathrm{EM}}$, while its $R^2_{\mathrm{design}}$ is lower by 0.00258.
The design-performance difference is concentrated in maximum gain, where the CNN and residual model achieve $R^2$ values of 0.86045 and 0.87938, respectively.
These results indicate that residual blocks are not essential to achieving high aggregate accuracy in this setting.

\section{Physical Consistency of Predicted EM Responses}
\label{app:passivity}

Prediction accuracy alone does not guarantee that a learned EM response satisfies physical constraints.
Our reduced S-parameter representation enforces reciprocity through reconstruction of the full matrix, but the model does not explicitly enforce passivity.
We therefore measure passivity violations in the predicted responses of both model sizes.

We sample 1,000 layouts without replacement from the test set and evaluate all 501 frequencies for each layout.
The same layouts are used for both model sizes and all three matching settings.
After denormalizing the predictions and reconstructing the full complex S-parameter matrix $S$, we count a violation when its largest singular value satisfies $\sigma_{\max}(S)>1+10^{-6}$.
The reported rate is the fraction of the resulting $N=501{,}000$ layout--frequency pairs that violate this criterion.

\begin{table}[htbp]
    \centering
    \caption{Passivity-violation rates in percent. Error bars are one IID Bernoulli standard error over layout--frequency pairs, computed as $100\sqrt{\hat p(1-\hat p)/N}$ for the observed violation fraction $\hat p$. This calculation treats pairs as independent and does not account for dependence among frequencies of the same layout.}
    \label{tab:passivity_violations}
    \begin{tabular}{lcc}
        \toprule
        Matching & 5.4M & 21M \\
        \midrule
        Input & $7.118 \pm 0.036$ & $3.925 \pm 0.027$ \\
        Interstage & $8.170 \pm 0.039$ & $4.134 \pm 0.028$ \\
        Output & $13.874 \pm 0.049$ & $6.890 \pm 0.036$ \\
        \bottomrule
    \end{tabular}
\end{table}

The 21M model exhibits fewer violations in every matching setting, with rates below 7\% across the evaluated layout--frequency pairs.
Neither model guarantees passivity.
For reference, the ground-truth responses on the same sample have a violation rate of 0.138\% under the same numerical criterion.
Thus, improved prediction accuracy accompanies fewer passivity violations, while explicit physical consistency remains a separate requirement.

\section{Solver-Stability Evaluation Protocol}
\label{app:solver_stability}
\label{app:noise_examples}

\Cref{fig:solver_instability} compares the robustness of the learned neural solver and the ground-truth (GT) solver when their input responses are imperfect.
The experiment tests how small response errors propagate to design-performance predictions in the input-matching setting.

\subsection{Noise Injection and Solver Evaluation}

For the GT solver, we add independent zero-mean Gaussian noise to every dimension of the ground-truth EM response.
We vary the noise standard deviation from 1\% to 20\% of the empirical standard deviation of the corresponding response dimension, in increments of one percentage point, and include a noiseless reference.
Noise is applied to the normalized magnitude and phase dimensions before denormalization and reconstruction of the complex S-parameter matrix.
We then feed this matrix to the GT solver, keeping frequency and circuit context unchanged.

For the learned solver, we perturb the predicted neural EM response using the same dimension-wise noise-scaling rule.
This input includes both the explicit EM response and the complementary response.
The trained solver remains fixed throughout the experiment, and frequency and circuit context remain unchanged.
We compare the resulting design-performance predictions from both solvers against the ground-truth design targets using $R^2_{\mathrm{design}}$. For this study, we compute each design-target score separately at each frequency over layouts and circuit contexts, then average over frequencies and targets. The study implementation uses $1-\mathrm{MSE}/s_y^2$, where $s_y^2$ is the sample variance with denominator $n-1$ for $n$ observations. The EM-response score uses the same variance convention, pooling layouts and frequencies within each response dimension before averaging over dimensions.

\subsection{Magnitude and Effect of EM Simulation Response Noise}

The GT solver's design-performance $R^2$ already becomes negative at 4\% noise.
To illustrate how much this noise level changes the simulated EM response, \cref{tab:noise_examples} shows the first complex S-parameter entry of three randomly sampled examples before and after perturbation.

\begin{table}[htbp]
    \centering
    \caption{Example S-parameter values before and after adding 4\% noise. Here $j$ denotes the imaginary unit.}
    \label{tab:noise_examples}
    \begin{tabular}{ccc}
        \toprule
        Example & Before & After \\
        \midrule
        1 & $-0.4457 + 0.2717j$ & $-0.4396 + 0.2774j$ \\
        2 & $-0.1701 + 0.3195j$ & $-0.1654 + 0.3267j$ \\
        3 & $-0.3637 + 0.3805j$ & $-0.3585 + 0.3726j$ \\
        \bottomrule
    \end{tabular}
\end{table}

These examples show that the added noise changes the simulated values only slightly, preserving the dominant EM signal.
The perturbed EM responses also retain an aggregate $R^2_{\mathrm{EM}}$ of 0.9984, confirming that the response remains close to the ground truth despite the GT solver's degraded performance.
In contrast, the learned solver retains $R^2_{\mathrm{design}}=0.9808$ at the same relative noise level, compared with 0.9814 without added noise.
Its accuracy decreases gradually as the noise magnitude increases.
We do not inject artificial response noise during training to enforce this stability.
The behavior emerges through joint training, possibly because the learned solver is exposed to imperfect EM-response predictions from the neural simulator.
The GT solver's sensitivity to these small EM-response errors motivates direct joint prediction of EM response and design performance.

\par
\clearpage
\endgroup

}{}

\clearpage
\section*{NeurIPS Paper Checklist}

\begin{enumerate}

\item {\bf Claims}
    \item[] Question: Do the main claims made in the abstract and introduction accurately reflect the paper's contributions and scope?
    \item[] Answer: \answerYes{}
    \item[] Justification: The introduction states the motivation, scope, contributions, model components, dataset, accuracy, and speed claims. These claims are supported by the method and experiment sections, with limitations summarized in \cref{sec:discussion}.

\item {\bf Limitations}
    \item[] Question: Does the paper discuss the limitations of the work performed by the authors?
    \item[] Answer: \answerYes{}
    \item[] Justification: Limitations are discussed in \cref{sec:discussion}, including the fixed-grid and process-specific setting, separate models for each matching placement, limited cross-distribution generalization, interpolation-only frequency evaluation, training cost, unquantified random-seed variability, and lack of guaranteed passivity.

\item {\bf Theory assumptions and proofs}
    \item[] Question: For each theoretical result, does the paper provide the full set of assumptions and a complete (and correct) proof?
    \item[] Answer: \answerNA{}
    \item[] Justification: The paper does not present formal theoretical results requiring assumptions and proofs. The mathematical expressions in \cref{sec:method} define the model objective and regularizer.

\item {\bf Experimental result reproducibility}
    \item[] Question: Does the paper fully disclose all the information needed to reproduce the main experimental results of the paper to the extent that it affects the main claims and/or conclusions of the paper (regardless of whether the code and data are provided or not)?
    \item[] Answer: \answerYes{}
    \item[] Justification: \Cref{sec:background,sec:method,sec:experiments} disclose the task, dataset construction, model architecture, objective, data split, optimization setup, training duration, and evaluation protocol. \Cref{app:backbone_comparison,app:passivity,app:solver_stability} provide the additional backbone, physical-consistency, and solver-perturbation evaluation protocols. The dataset cannot be released due to commercial licensing constraints, but the paper describes the dataset construction details needed for reproduction, and the provided code supports reproducing the main experimental results supporting the paper's claims.

\item {\bf Open access to data and code}
    \item[] Question: Does the paper provide open access to the data and code, with sufficient instructions to faithfully reproduce the main experimental results, as described in supplemental material?
    \item[] Answer: \answerYes{}
    \item[] Justification: The code is provided, but the dataset cannot be publicly released due to commercial licensing constraints. To support reproducibility despite this restriction, the paper documents the dataset construction, model, training setup, and evaluation protocol in \cref{sec:method,sec:experiments}, providing the details needed to recreate the dataset and reproduce the reported results.

\item {\bf Experimental setting/details}
    \item[] Question: Does the paper specify all the training and test details (e.g., data splits, hyperparameters, how they were chosen, type of optimizer) necessary to understand the results?
    \item[] Answer: \answerYes{}
    \item[] Justification: \Cref{sec:experiments} specifies the data split, target and conditioning normalization, augmentation, model scales, optimizer, learning-rate schedule, batch size, checkpoint selection, training duration, and hyperparameter-selection protocol. Additional evaluation settings are described in \cref{app:backbone_comparison,app:passivity,app:solver_stability}.

\item {\bf Experiment statistical significance}
    \item[] Question: Does the paper report error bars suitably and correctly defined or other appropriate information about the statistical significance of the experiments?
    \item[] Answer: \answerNo{}
    \item[] Justification: The main predictive-performance results are from single training runs, and variability across random seeds is not quantified, as stated in \cref{sec:experiments,sec:discussion}. \Cref{app:passivity} reports one IID Bernoulli standard error for passivity-violation rates and explicitly states the independence assumption. These error bars do not measure training-seed variability or account for dependence among frequencies of the same layout.

\item {\bf Experiments compute resources}
    \item[] Question: For each experiment, does the paper provide sufficient information on the computer resources (type of compute workers, memory, time of execution) needed to reproduce the experiments?
    \item[] Answer: \answerYes{}
    \item[] Justification: \Cref{sec:experiments} reports the main hardware setup, including 8 NVIDIA GeForce RTX 2080 Ti GPUs, per-GPU batch size, storage footprint, and approximate training time for the 5.4M and 21M models. These details specify the resources used for the main experiments.
    
\item {\bf Code of ethics}
    \item[] Question: Does the research conducted in the paper conform, in every respect, with the NeurIPS Code of Ethics \url{https://neurips.cc/public/EthicsGuidelines}?
    \item[] Answer: \answerYes{}
    \item[] Justification: The work uses simulated RFIC passive-layout data and neural surrogate modeling. We are not aware of any deviation from the NeurIPS Code of Ethics.

\item {\bf Broader impacts}
    \item[] Question: Does the paper discuss both potential positive societal impacts and negative societal impacts of the work performed?
    \item[] Answer: \answerNA{}
    \item[] Justification: The paper focuses on technical modeling of RFIC passives and does not include a dedicated broader-impact discussion. The main expected impact is faster RFIC passive design, with limited direct societal-risk pathways.
    
\item {\bf Safeguards}
    \item[] Question: Does the paper describe safeguards that have been put in place for responsible release of data or models that have a high risk for misuse (e.g., pre-trained language models, image generators, or scraped datasets)?
    \item[] Answer: \answerNA{}
    \item[] Justification: The released asset is code, not a high-risk model, scraped dataset, human-subject dataset, or public release of the RFIC mask dataset requiring special misuse safeguards.

\item {\bf Licenses for existing assets}
    \item[] Question: Are the creators or original owners of assets (e.g., code, data, models), used in the paper, properly credited and are the license and terms of use explicitly mentioned and properly respected?
    \item[] Answer: \answerNA{}
    \item[] Justification: The work does not use existing external datasets or pretrained models as research assets. Relevant prior methods and software components are cited where they are discussed.

\item {\bf New assets}
    \item[] Question: Are new assets introduced in the paper well documented and is the documentation provided alongside the assets?
    \item[] Answer: \answerYes{}
    \item[] Justification: The paper provides code to support reproducibility and future research. The dataset construction, model setup, and experimental details are documented in \cref{sec:method,sec:experiments}.

\item {\bf Crowdsourcing and research with human subjects}
    \item[] Question: For crowdsourcing experiments and research with human subjects, does the paper include the full text of instructions given to participants and screenshots, if applicable, as well as details about compensation (if any)? 
    \item[] Answer: \answerNA{}
    \item[] Justification: The paper does not involve crowdsourcing or human-subject experiments.

\item {\bf Institutional review board (IRB) approvals or equivalent for research with human subjects}
    \item[] Question: Does the paper describe potential risks incurred by study participants, whether such risks were disclosed to the subjects, and whether Institutional Review Board (IRB) approvals (or an equivalent approval/review based on the requirements of your country or institution) were obtained?
    \item[] Answer: \answerNA{}
    \item[] Justification: The paper does not involve crowdsourcing or human-subject experiments, so IRB approval is not applicable.

\item {\bf Declaration of LLM usage}
    \item[] Question: Does the paper describe the usage of LLMs if it is an important, original, or non-standard component of the core methods in this research? Note that if the LLM is used only for writing, editing, or formatting purposes and does \emph{not} impact the core methodology, scientific rigor, or originality of the research, declaration is not required.
    \item[] Answer: \answerYes{}
    \item[] Justification: LLM assistance is disclosed in the unnumbered LLM Disclosure section immediately before the references. The assistance included brainstorming, partial implementation assistance, and writing polish. The authors remain fully responsible for the final content.

\end{enumerate}

\message{^^JLASTPAGE \thepage^^J}

\end{document}